\documentclass[aps,prd,onecolumn,groupedaddress,showpacs,nofootinbib,amssymb]{revtex4}
\usepackage[pdftex]{graphicx}
\usepackage{amssymb}
\usepackage{amsmath}
\usepackage{graphicx}
\usepackage{amsfonts}
\usepackage{bm}
\usepackage{latexsym,float}
\usepackage{epsfig}
\usepackage{epstopdf}
\usepackage{amssymb}
\usepackage{verbatim}
\usepackage{multirow}
\usepackage{color}

\begin{document}

\title{Bianchi I cosmological solutions in teleparallel gravity}



\author{Petr V. Tretyakov}
\email{tpv@theor.jinr.ru} \affiliation{Joint Institute for
Nuclear Research, Joliot-Curie 6, 141980 Dubna, Moscow region,
Russia}

\begin{abstract}
In the present paper, we discuss the dynamics of the an\-iso\-trop\-ic Bianchi I Universe filled with perfect fluid in teleparallel $f(T)$-gravity. Using both analytical and numerical approaches, we confirm the main results of previous authors such as the presence of a future isotropic attractor, and find a number of new ones such as the existence of an essential number of bounce solutions and the possibility to construct late time acceleration without any specific matter.
\end{abstract}


\maketitle

\section{Introduction}\label{sec:1}

Teleparallel gravity is an alternative way to describe the gravitational field\cite{Einstein,AP,JRSV}. It uses the so-called Weitzenb\"{o}ck connection instead of the Levi-Civita one used in Riemann geometry. The main objects in this formulation of gravity are the tetrad field $e_A(x^{\mu})$, spin connection $\omega^{A}_{\,\,\,\,B\mu}$, which is equal to zero for a set of ansatz (e.g., the description of the FRW Universe in the cartesian coordinates), and the torsion scalar $T$ that replaces the curvature scalar $R$ in action. Note that in the simplest formulation teleparallel gravity has well-known problems with local Lorentz invariance\cite{LSB,SLB,TB}, and the introduction of spin connection is one of the ways\footnote{The other way is to use proper tetrads.} to solve these problems\cite{KS2}. The formulation of teleparallel gravity with taking into account spin connection is also known as the covariant formulation of teleparallel gravity\cite{G2,HJU,G3}. We will not use spin connection in the present research, because our ansatz implies its zero value without loss of generality. It is well known that in the linear case $f(T)=T$, this formulation of gravity is totally identical to the General Relativity (GR), but for non-trivial functions this approach gives essentially different results in comparison with the $f(R)$-theories \cite{Linder}, for instance, $f(R)$-gravity produces fourth-order equations, whereas equations for $f(T)$-gravity have second order, as in GR. For this reason, teleparallel gravity has become an area of intense investigations\cite{KS1,Tret, arg1,arg2,BMNO,BCNO,BOSG,BCLNSG,N1,CCLS,N2,CLSX,WG,G1,TT,SB1}.

From general considerations there are no arguments why the initial singularity should be isotropic and the study of anisotropic spaces at early times was always an important task. The most known anisotropic solution in GR is Kasner's solution, which is the vacuum solution for Bianchi I geometry. Bianchi I space is the simplest case of anisotropic geometry and therefor it attracts the attention of researchers, who study different modified gravity theories and, in particular, the $f(T)$-gravity models\cite{SS1,RHSGR,SA,N3,SS2,PBL,WNI,PSB}. The pure Kasner solution in Bianchi I $f(T)$ cosmology was studied in \cite{ST1}. A quite successful attempt to study Bianchi I ansatz in teleparallel $f(T)$-gravity with perfect fluid was made in \cite{ST2}, where the authors by using dynamical system approach and dimensionless variables found a set of very interesting results. First of all, they found that the presence of matter plays the key role in anisotropic dynamics near a cosmological
singularity. Moreover, they found de Sitter points and future isotropic attractor and computed a set of cosmological scenarios numerically. Nevertheless, the authors missed some very interesting and important things due to the limited scope of their approach.

In the present paper, we reproduce the main results from \cite{ST2} by using some different approaches and essentially complement them. This paper is organized as follows. In Sec.2, we briefly remind the key points of the teleparallel gravity formulation. In Sec.3, we derive equations of motion. In Sec.4, we demonstrate the existence of a future isotropic attractor, analyze its stability and describe all possible stationary points for our system. In Sec.5, we supplement our results with numerical investigations, and in Sec.6, we give our concluding remarks.

\section{Formulation of teleparallel gravity}\label{sec:2}

In the standard formulation of teleparallel gravity the main geometrical object is the so-called tetrad field
\begin{equation}
e_A(x^{\mu}),
\label{0.1}
\end{equation}
 defined  in tangent space. The space-time metric tensor can be expressed through the tetrad field as
\begin{equation}
g_{\mu\nu}=\eta_{AB}e^A_{\,\,\,\,\mu}e^B_{\,\,\,\,\nu},
\label{0.2}
\end{equation}
where $\eta_{AB}=\mathrm{diag}(1,-1,-1,-1)$ is the Minkowski matrix.
Here and below the capital Latin indices refer to the tangent space and take values $0..3$,  whereas the Greek indices refer to space-time and also take values $0..3$.
The torsion tensor is defined as
\begin{equation}
T^{\lambda}_{\,\,\,\,\mu\nu}\equiv \Gamma^{\lambda}_{\,\,\,\,\nu\mu} -\Gamma^{\lambda}_{\,\,\,\,\mu\nu} = e_A^{\,\,\,\,\lambda}(\partial_{\mu}e^A_{\,\,\,\,\nu}-\partial_{\nu}e^A_{\,\,\,\,\mu}),
\label{0.3}
\end{equation}
where $e_A^{\,\,\,\,\mu}$ denotes the inverse tetrad, which satisfies $e_A^{\,\,\,\,\mu}e^A_{\,\,\,\,\nu}=\delta^{\mu}_{\nu}$ and $e_A^{\,\,\,\,\mu}e^B_{\,\,\,\,\mu}=\delta^{B}_{A}$. In addition, the following tensors are usually defined: contorsion tensor, which equals the difference between the Weitzenb\"{o}ck and Levi-Civita connection
\begin{equation}
K^{\mu\nu}_{\,\,\,\,\,\,\,\,\lambda}\equiv -\frac{1}{2}\left ( T^{\mu\nu}_{\,\,\,\,\,\,\,\,\lambda} -T^{\nu\mu}_{\,\,\,\,\,\,\,\,\lambda} -T_{\lambda}^{\,\,\,\,\mu\nu} \right ),
\label{0.4}
\end{equation}
and the following auxiliary tensor:
\begin{equation}
S_{\lambda}^{\,\,\,\,\mu\nu}\equiv (K^{\mu\nu}_{\,\,\,\,\,\,\,\,\lambda}+\delta^{\mu}_{\lambda} T^{\alpha\nu}_{\,\,\,\,\,\,\,\,\alpha} -\delta^{\nu}_{\lambda} T^{\alpha\mu}_{\,\,\,\,\,\,\,\,\alpha})
\label{0.5}
\end{equation}

Note that the last tensor and torsion tensor are antisymmetric
at the second and third indices, so\footnote{Note here that all Greek indices are raised and lowered by using the metric; whereas the Latin indices, by using the Minkowski matrix.} $T_{\lambda(\mu\nu)}=S_{\lambda(\mu\nu)}=0$, whereas for the contorsion tensor we have $K_{(\mu\nu)\lambda}=0$.

Now the torsion scalar can
be defined as
\begin{equation}
T\equiv \frac{1}{2} S_{\lambda}^{\,\,\,\,\mu\nu}T^{\lambda}_{\,\,\,\,\mu\nu}  =\frac{1}{4}T^{\lambda\mu\nu}T_{\lambda\mu\nu}+\frac{1}{2}T^{\lambda\mu\nu}T_{\nu\mu\lambda}- T_{\lambda\mu}^{\,\,\,\,\,\,\,\,\lambda}T^{\nu\mu}_{\,\,\,\,\,\,\,\,\nu},
\label{0.6}
\end{equation}
and the action of teleparallel gravity in the most general form can be written as
\begin{equation}
S=\frac{1}{2\kappa^2}\int d^4x\, e\, f(T),
\label{0.7}
\end{equation}
where $e=det(e^A_{\,\,\,\,\mu})=\sqrt{-g}$ and $\kappa^2$ is the gravitational constant. Variation of action (\ref{0.7}) with respect to the tetrad field gives the equation of motion in the following form:
\begin{equation}
e^{-1}\partial_{\mu}(eS_A^{\,\,\,\,\nu\mu})f'-e^{\,\,\,\,\lambda}_AT^{\rho}_{\,\,\,\,\mu\lambda}S_{\rho}^{\,\,\,\,\mu\nu}f' + S_A^{\,\,\,\,\mu\nu}\partial_{\mu}(T)f''+\frac{1}{2}e^{\,\,\,\,\nu}_Af=\kappa^2 e^{\,\,\,\,\rho}_A T_{\,\,\,\rho}^{m\,\,\,\,\nu},
\label{0.8}
\end{equation}
which is also can be rewritten in equivalent form \cite{LSB1}
\begin{equation}
\tilde{G}_{\mu\nu}\equiv f'(\stackrel{\!\!\!\!\!\!\circ}{R_{\mu\nu}}-\frac{1}{2}g_{\mu\nu}\stackrel{\circ}{R}) + \frac{1}{2}g_{\mu\nu}[f(T)-f'T] +f''S_{\nu\mu\lambda}\nabla^{\lambda}T=\kappa^2 T^{m}_{\mu\nu},
\label{0.9}
\end{equation}
where $\stackrel{\!\!\!\!\!\!\circ}{G_{\mu\nu}}\equiv \stackrel{\!\!\!\!\!\!\circ}{R_{\mu\nu}}-\frac{1}{2}g_{\mu\nu}\stackrel{\circ}{R}$ should
be calculated by using the Levi-Civita connection $\stackrel{\!\!\!\!\!\!\!\!\!\!\!\circ}{\Gamma^{\rho}_{\,\,\,\,\mu\nu}}=\frac{1}{2}g^{\rho\sigma}(\partial_{\mu}g_{\sigma\nu}+\partial_{\nu}g_{\mu\sigma} -\partial_{\sigma}g_{\mu\nu})$.

This formulation of teleparallel gravity (without taking into account spin connection) has the well-known problem of lack of local Lorentz invariance.
There are two ways to fix this problem: one using spin connection and the other using the proper  tetrad.  
 The proper tetrad we need is usually {\it a priory} unknown. Nevertheless, as we already mentioned, for our case we can use this formulation of teleparallel gravity because the spin connection is equal to zero for the flat Bianchi I ansatz without perturbations.

\section{Equations}\label{sec:3}

We start with the following tetrad, which corresponds to the Bianchi I geometry:

\begin{equation}
e^A_{\mu}=\mathrm{diag}\big (1,\mathrm{a}(t),\mathrm{b}(t),\mathrm{c}(t)\big ),
\label{1.1}
\end{equation}
and the corresponding metric is
\begin{equation}
g_{\mu\nu}=\mathrm{diag}\big (1,-\mathrm{a}(t)^2,-\mathrm{b}(t)^2,-\mathrm{c}(t)^2\big ).
\label{1.2}
\end{equation}
Now introducing three Hubble parameters $H_{\mathrm{a}}\equiv\frac{\mathrm{\dot a}}{\mathrm{a}}$, $H_{\mathrm{b}}\equiv\frac{\mathrm{\dot b}}{\mathrm{b}}$ and $H_{\mathrm{c}}\equiv\frac{\mathrm{\dot c}}{\mathrm{c}}$, we find for torsion scalar
\begin{equation}
T=-2\big( H_\mathrm{a}H_\mathrm{b}+H_\mathrm{a}H_\mathrm{c}+H_\mathrm{b}H_\mathrm{c} \big).
\label{1.3}
\end{equation}
Now the $00$-component of equation (\ref{0.9}) reads

\begin{equation}
\frac{1}{2}f-Tf'=\kappa^2\rho,
\label{1.4}
\end{equation}
where we denote $f'\equiv df/dT$ and introduce isotropic fluid in the right-hand side $p=w\rho$, the conservation equation for which reads
\begin{equation}
\dot\rho+\big( 1+w\big )\big( H_\mathrm{a}+H_\mathrm{b}+H_\mathrm{c}\big )\rho=0.
\label{1.5}
\end{equation}
Finally, non-trivial spatial (diagonal) components of equation (\ref{0.9}) take the following form
\begin{eqnarray}
\big(H_\mathrm{b}+H_\mathrm{c}\big)\dot T f''+\frac{1}{2}f +f'\big(\dot H_\mathrm{b}+H_\mathrm{b}^2+\dot H_\mathrm{c}+H_\mathrm{c}^2+2H_\mathrm{b}H_\mathrm{c}+H_\mathrm{a}H_\mathrm{b}+H_\mathrm{a}H_\mathrm{c} \big)=&&-\kappa^2w\rho,\label{1.6}\\
\big(H_\mathrm{a}+H_\mathrm{c}\big)\dot T f''+\frac{1}{2}f +f'\big(\dot H_\mathrm{a}+H_\mathrm{a}^2+\dot H_\mathrm{c}+H_\mathrm{c}^2+2H_\mathrm{a}H_\mathrm{c}+H_\mathrm{a}H_\mathrm{b}+H_\mathrm{b}H_\mathrm{c} \big)=&&-\kappa^2w\rho,\label{1.7}\\
\big(H_\mathrm{a}+H_\mathrm{b}\big)\dot T f''+\frac{1}{2}f +f'\big(\dot H_\mathrm{a}+H_\mathrm{a}^2+\dot H_\mathrm{b}+H_\mathrm{b}^2+2H_\mathrm{a}H_\mathrm{b}+H_\mathrm{a}H_\mathrm{c}+H_\mathrm{b}H_\mathrm{c} \big)=&&-\kappa^2w\rho.\label{1.8}
\end{eqnarray}
Note that equation (\ref{1.4}) is the first integral of system (\ref{1.5}-\ref{1.8}). For equations (\ref{1.6}-\ref{1.8}) we can see the following situation: three different functions (left-hand sides) are equal to the absolutely identical value (r.h.s.). This fact tells us that actually the number of independent variables is less. This kind of picture for Bianchi I is well known for us in the GR and $f(R)$-gravity. To derive a reduced system, we introduce a new (quite natural) variable

\begin{equation}
H\equiv H_\mathrm{a}+H_\mathrm{b}+H_\mathrm{c},
\label{1.9}
\end{equation}
and try to derive a system for variables\footnote{Note here that since our variable $H$ is not an average value of three Hubble parameters but the sum, for the isotropic limit ($H_\mathrm{a}=H_\mathrm{b}=H_\mathrm{c}$) we have the relation $T=-2H^2/3=-6H_\mathrm{a}^2$. } $H$ and $T$. The first equation can be derived very easily by summing equations (\ref{1.6}) $+$ (\ref{1.7}) $+$ (\ref{1.8}):
\begin{equation}
2\dot THf''+\frac{3}{2}f+2f'(\dot H+H^2)=-3w\kappa^2\rho,
\label{1.10}
\end{equation}
and the second one by the summing with the coefficients $(-2H_a)\cdot$ (\ref{1.6}) $ +(-2H_b)\cdot$ (\ref{1.7}) $+(-2H_c)\cdot$ (\ref{1.8}):
\begin{equation}
2\dot TTf''-Hf +\dot Tf'+2f'TH=2w\kappa^2H\rho,
\label{1.11}
\end{equation}
and instead of (\ref{1.5}) we have obviously
\begin{equation}
\dot\rho+\big( 1+w\big )H\rho=0.
\label{1.12}
\end{equation}

Now we have the system of three equations (\ref{1.10}-\ref{1.12}) for three {\it independent} variables $H$, $T$, $\rho$ with the first integral (\ref{1.4}). It is interesting to note that the variable $\rho$ may be totally decoupled from the system: if we derive the matter density from the constraint (\ref{1.4}) and substitute it into equations (\ref{1.10})-(\ref{1.11}), we obtain a closed system of two differential equations for two variables $H$ and $T$,  and the only  influence of matter density will by the parameter $w$ of the equation of state.

Let us note once again that the reduction from the $4$d system (\ref{1.5})-(\ref{1.8}) to the $3$d system (\ref{1.10})-(\ref{1.12}) is possible due to the existence of hidden symmetry in the Bianchi I space. To produce the back transformation from our variables ($H$, $T$) to the initial ($H_a$, $H_b$, $H_c$), we must use expressions (\ref{1.3}), (\ref{1.9}) and one additional expression that can be derived by using the initial equations (\ref{1.6})-(\ref{1.8}). We do not need this relation in our paper, but it was derived by the previous authors in \cite{ST2} and reads $H_a=(1+C)H_b-CH_c$, where $C$ is integration constant.

\section{Future attractor and stationary points}\label{sec:4}

For the expanding Universe we expect decreases of all dynamical variables with time, so let us try to find solution in the form $T=\mathrm{t_0}t^{-m}$, $H=h_0t^{-n}$, $\rho=\rho_0t^{-k}$ with some positive $m,\,n,\,k$. Substituting it into (\ref{1.12}), we find $n=1$, which is quite natural. Now we can see that for any function $f$ that can be expanded in the Taylor series $f=\sum_{i=1}^{\infty}f_iT^i$ will keep only the terms with the lowest $i$ because all other will decrease more rapidly. It is quite natural to suppose that the lowest term is $T$ because $f=T$ is equivalent to GR, and we would like to have it as a limit. It means that instead of (\ref{1.10})-(\ref{1.11}) for this kind of solution we have an approximate system
\begin{eqnarray}
\frac{3}{2}T+2\big( \dot H+H^2\big)=&&-3\kappa^2w\rho=-3w(\frac{1}{2}T-T),\label{1.13}\\
H T+\dot T=&&2\kappa^2wH\rho=2wH(\frac{1}{2}T-T),\label{1.14}
\end{eqnarray}
where we used constraint (\ref{1.4}). Substituting there our solution, we find $m=k=2$. Let us introduce the parameter $a$: $T=-aH^2$. Now for the parameters $a$, $h_0$ we have
\begin{eqnarray}
-\frac{3}{2}ah_0^2-2h_0+2h_0^2=&&-\frac{3}{2}wh_0^2a,\label{1.15}\\
-h_0ah_0^2+2ah_0^2=&&wh_0ah_0^2,\label{1.16}
\end{eqnarray}
which has the unique solution $h_0(1+w)=2$, $a=2/3$ corresponding to the expanding ($h_0>0,\,\forall w$) isotropic solution. It should be noted that this solution exists for any $w$ and any function $f$ (which contain GR as a limit), but the stability of this solution will obviously depend on a concrete shape of $f$ and (maybe) on $w$.

\subsection{Stability of obtained solutions}

Let us discuss the stability of arbitrary solution. For this we reduce the system to two dimensional system by using the constraint equation. We have

\begin{eqnarray}
2\dot T H f'' +\frac{3}{2}(1+w)f -3wTf' + 2f'(\dot H +H^2)=&&0,\label{1.17}\\
2\dot T T f'' -(1+w)Hf +\dot T f' + 2 (1+w) HTf' =&&0.\label{1.18}
\end{eqnarray}
Let us suppose that we have some solution of this system $T_0=T_0(t)$, $H_0=H_0(t)$ and small perturbations near this solution $\delta(t)$, $\gamma(t)$ such that $T=T_0+\delta$, $H=H_0+\gamma$. Expanding all functions in the system to the Taylor series by the scheme $f=f_0+f_0'\delta$, where $f_0\equiv f(T_0)$ and keeping only linear terms with respect to perturbations, we get a general system that governs the stability of the discussed solution:

\begin{eqnarray}
2H_0f_0''\dot\delta+2\dot T_0f_0''\gamma+2\dot T_0H_0f_0'''\delta+\frac{3}{2}(1-w)f_0'\delta-3wT_0f_0''\delta+2(\dot H_0+H_0^2)f_0''\delta+2f_0'(\dot\gamma+2H_0\gamma)=&&0,\label{1.19}\\
\!\!\!\!\!\!\!\!\!\!\!\!2T_0f_0''\dot\delta +3\dot T_0f_0'' \delta +2\dot T_0T_0f_0'''\delta -(1+w)f_0\gamma +f_0'\dot\delta +2(1+w)T_0f_0'\gamma +(1+w)H_0f_0'\delta +2(1+w)H_0T_0f_0''\delta =&&0.\label{1.20}
\end{eqnarray}

We can see that in the most general case the system looks like extremely complicate; thus let try to estimate the stability of a future attractor for the simplest case of the function $f=T+f_2T^2$. It is quite natural expect a power law solution for a future attractor. Thus substituting $H=h_0t^n$ into the equation (\ref{1.12}), we easily get $n=-1$ and, therefore, the solution for a future attractor reads $H_0=h_0t^{-1}$, $T_0=-2h_0^2/(3t^2)$. Now substituting this solution into equations (\ref{1.19})-(\ref{1.20}) and taking into account that if a decreasing mode exists, we have $t^{-2}\delta\ll t^{-1}\delta$ and so on. We find two approximate equations:
\begin{eqnarray}
\dot\delta-\frac{2}{3}h_0^2(1+w)t^{-2}\gamma + h_0(1+w)t^{-1}\delta=&&0,\label{1.21}\\
4f_2h_0t^{-1}\dot\delta +\frac{3}{2}(1-w)\delta + 2\dot\gamma +4h_0t^{-1}\gamma =&&0.\label{1.22}
\end{eqnarray}
This system has an exact solution but it is not demonstrative; so let us try to analyze it in different limits to understand whether it can have a decreasing mode or not.

First of all let us put $\gamma=0$. The solution of equation (\ref{1.21}) reads $\delta\propto t^{-h_0(1+w)}$ and the solution of equation (\ref{1.22}) reads  $\delta\propto t^{-\frac{3(1-w)t^2}{8f_2h_0}}$. Both are decreasing modes if $f_2>0$ for any matter. Now let us put $\delta=0$; in this case, we are forced to ignore equation (\ref{1.21}) because it does not contain any dynamics and equation (\ref{1.22}) gives us $\gamma\propto t^{-2h_0}$ which is the decreasing mode. Now let us put $\gamma=\delta$. In this case, the exact solution of equation (\ref{1.21}) reads $\delta\propto t^{-h_0(1+w)}\mathrm{e}^{-\frac{2h_0^2(1+w)}{3t}}$ and for (\ref{1.22}) the exact solution reads $\delta\propto (4f_2h_0+2t)^{\frac{3}{2}(1-w)f_2h_0-2h_0}\mathrm{e}^{-\frac{3}{4}(1-w)t}$, both are the decreasing modes.

Surely, this approximate analysis cannot confirm that only decreasing modes of perturbations can exist, but it demonstrate that in the simplest cases decreasing modes can exist and therefor we may hope that this future attractor will be stable. A more general analysis of the stability of this attractor we address to the next section using a numerical approach.

\subsection{Stationary points}\label{sec:5}

Before we pass to numerical investigation of the model, let us try to understand the expected results. For this let us turn to the system of equations in the form of a two dimensional system (\ref{1.17})-(\ref{1.18}). This is a dynamical system, the stationary points of which are governed by the following equations:
\begin{eqnarray}
 +\frac{3}{2}(1+w)f -3wT_0f' + 2f'H_0^2=&&0,\label{1.26}\\
 -(1+w)H_0f + 2 (1+w) H_0T_0f' =&&0.\label{1.27}
\end{eqnarray}

In the most general case there are three (groups) stationary points.
\vspace{0.5cm}

${\bf P1}.$ The first one is $H_0=0$, $T_0=0$. This point exists for any shape of the function with $f(0)=0$, which is quite natural, any eos $w$ and corresponds to the Minkowski solution.
\vspace{0.5cm}

${\bf P2}.$ The second point is $H_0=0$, $T_0\neq 0$. $T_0:\,\,2wT_0f'=(1+w)f$. Note that this point exists only if the last equation admits solution with $T_0<0$ and can present a group of points for the polynomial function $f$.
\vspace{0.5cm}

${\bf P3}.$ And the third one is $H_0\neq 0$, $T_0\neq 0$. As we can see from (\ref{1.27}), in this case we have $T_0:\,\,2T_0f'=f$, $\forall w$ and therefore by using (\ref{1.26}), we find $2H_0^2=-3T_0$ that corresponds to the isotropic de Sitter solution. We can see that this solution actually corresponds to two de Sitter points one for an expanding and the other for a contracting Universe, and it exists only for functions that admit solutions with $T_0<0$. Additional analytical investigations for the function $f=T+f_2T^2$ demonstrate that i) it exists only for $f_2>0$, ii) it is always stable in the expanding Universe and unstable in the contracting one, iii) for reasonable values of $f_2<1$ it always lies in the non-physical region under the Planck scale.

\section{Numerical investigations}\label{sec:6}

We perform numerical investigations using two different approaches: the first, numerical integration with the use of the fourth order Runge-Kutta algorithm and the second, building of phase portraits of the system using a computer algebra program. The combination of both methods allows us to understand the real dynamical picture.

\subsection{Numerical integration}

For numerical integration we used the system (\ref{1.10})-(\ref{1.12}). This is the ODE system for three variables, thus we need three initial conditions for the variables $H$, $T$ and $\rho$. Note that only two of these initial values are independent because there exists a constraint equation (\ref{1.4}), which must be satisfied. Thus, we fixed the initial conditions for $H$ and $T$, whereas $\rho_0$ was calculated by using a constraint. Moreover, the existence of a constraint equation allow us to control the accuracy of the numerical solution and fix the point where numerical integration is diverges\footnote{It is possible to reduce the system to two equations by using a constraint, but in this case, we have no possibility to control the accuracy of the numerical solution.}.

First of all note that we need a physical interpretation of the contraction and expansion of the Universe in our case, because the variable $H$ is the sum of three Hubble parameters and the Universe can expand along one of the axes and contract on the others. Nevertheless, the existence of matter significantly simplifies  this task. Indeed, from the conservation equation (\ref{1.12}) it is quite clear that the point $H=0$ corresponds to the point where $\dot\rho=0$, i.e. the growth of the matter density changes by drop or contrary; thus we can interpret the point $H=0$ as a point of bounce. So we can interpret the regimes with $H<0$ as contracting Universe and the regimes with $H>0$ as expanding one. Moreover, if we substitute $H=0$ into equation (\ref{1.11}), we will find that at the point of bounce $\dot T=0$ as well.

\begin{figure}[h]
  \includegraphics[width=0.95\linewidth]{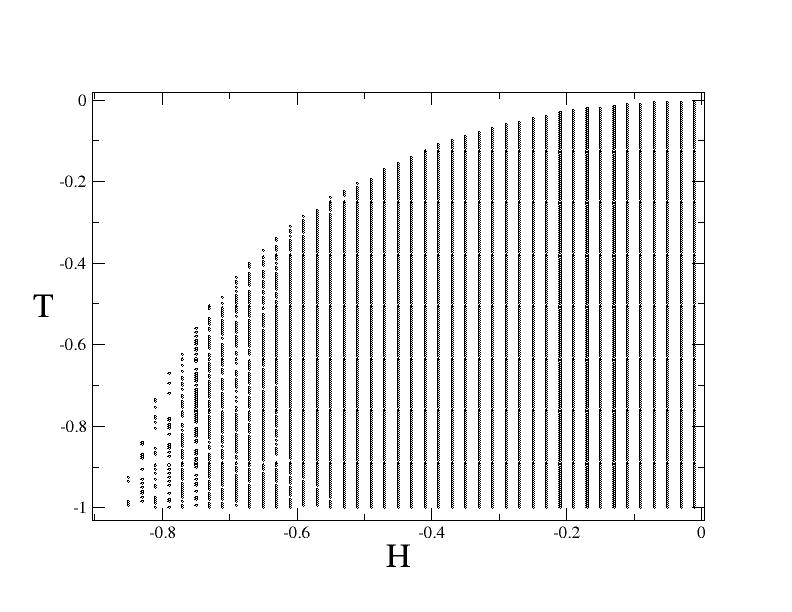}
  \caption{Bounce solutions for $f_2=0.1$, $N=2$, $w=0$, $\kappa^2=1$.}
  \label{fig:fig1}
\end{figure}

\begin{figure}[h]
  \includegraphics[width=0.95\linewidth]{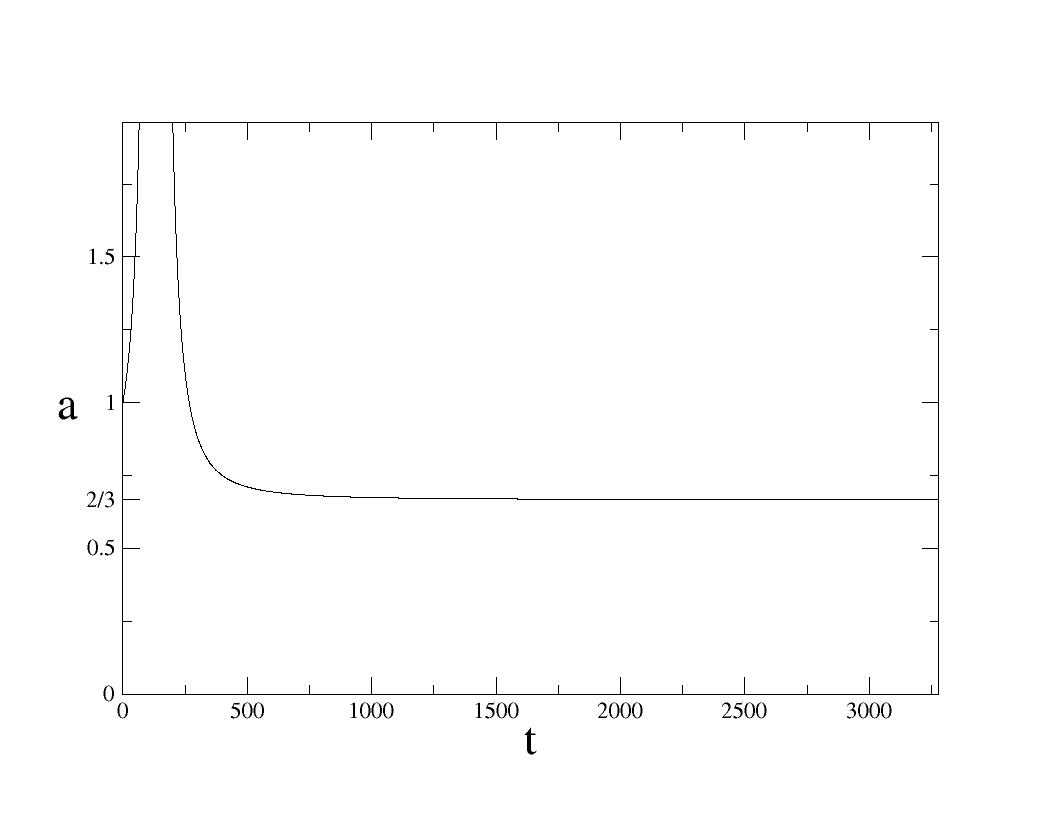}
  \caption{Evolution of the $a$-parameter for $f_2=0.1$, $N=2$, $w=0$, $\kappa^2=1$, $H_0=-10^{-2}$, $T_0=-10^{-4}$. At the bounce point the values of the parameter tend to infinity, after that quick isotropisation with $a=\frac{2}{3}$ occurs.}
  \label{fig:fig2}
\end{figure}

\begin{figure}[h]
  \includegraphics[width=0.95\linewidth]{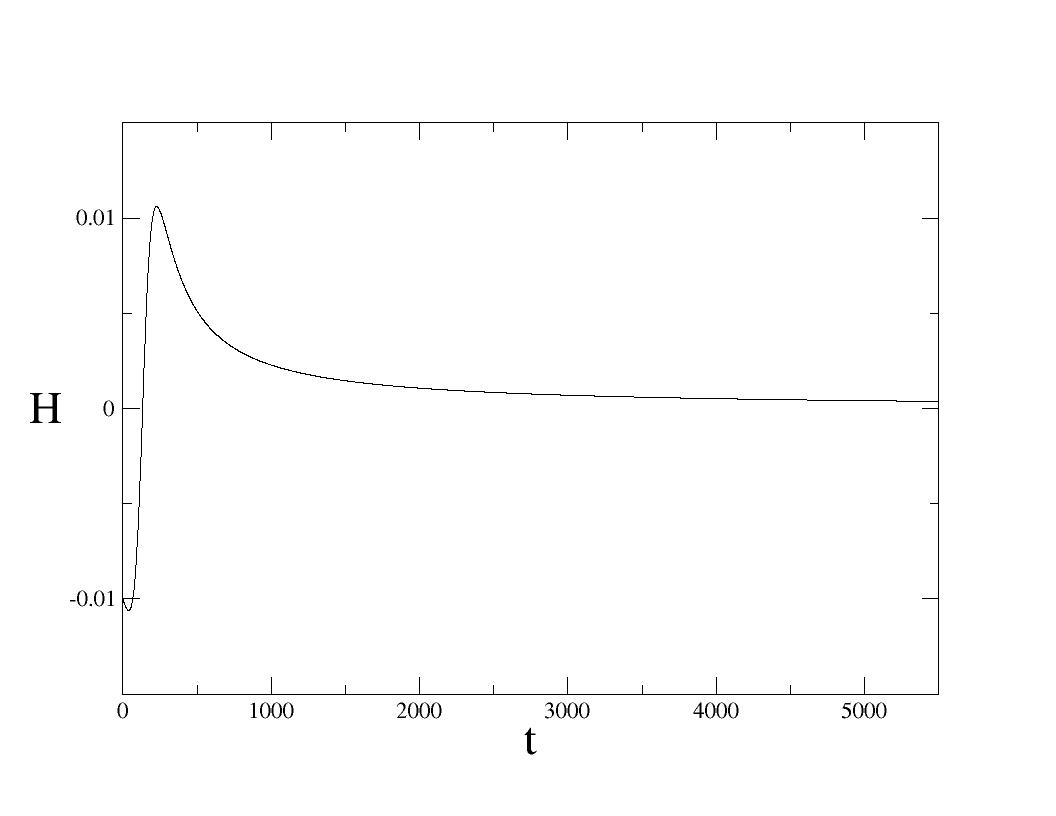}
  \caption{Evolution for an average Hubble parameter $H$ for $f_2=0.1$, $N=2$, $w=0$, $\kappa^2=1$, $H_0=-10^{-2}$, $T_0=-10^{-4}$. At the bounce point $H$ change the sign and tends to zero as time tends to infinity.}
  \label{fig:fig3}
\end{figure}

We found numerically such a kind of trajectories where contraction is changed by expansion and tends to an isotropic attractor. We study the function $f=T+f_2T^2$ and a typical number of bounce solutions is presented in fig.\ref{fig:fig1}. We studied different sets of parameters such as $f_2=\pm 0.1$, $w=\pm\frac{2}{3},\,\pm\frac{1}{3},\,0$ in all possible compositions and find that a number of bounce solutions is very essential for every set but slightly decreases for the hard matter ($w\rightarrow 1$). Not quite clear boundary appears due to some numerical instability near the separation line, it absolutely disappears for negative values of $f_0$ and $w\leqslant 0$. Typical behavior of the parameters $a$ and $H$ is presented in fig.\ref{fig:fig2} and fig.\ref{fig:fig3}, respectively.

\begin{figure}[h]
  \includegraphics[width=0.95\linewidth]{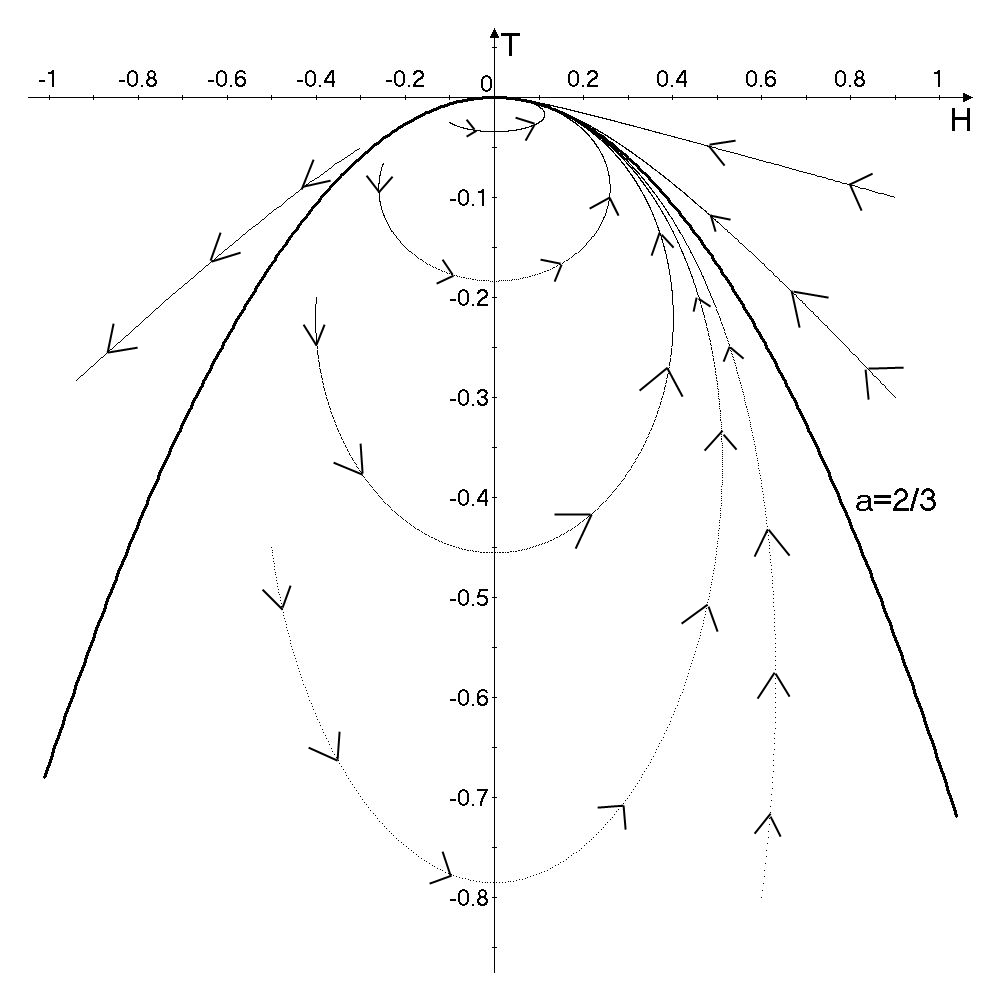}
  \caption{Future isotropic attractor for $f_2=0.1$, $N=2$, $w=0$, $\kappa^2=1$, $H_0=-10^{-2}$, $T_0=-10^{-4}$.}
  \label{fig:fig4}
\end{figure}

Another important thing, which was detected during analytical investigations, is a future isotropic attractor. We expect, according to approximate investigations, that it will be stable. Indeed, we integrate all found bounce solutions up to isotropisation at the level $|a-\frac{2}{3}|<10^{-4}$ and find that for any sets of parameters isotropisation is reached quickly. In fig.\ref{fig:fig4}, we present a numerical phase portrait for some concrete set of parameters in the physical region. This picture is quite typical for any power function. We can see a small number of trajectories that lead to a physical singularity during contraction ($H<0$), trajectories with bounce solution (that cross the zero line $H=0$ from left to right) and a future stable isotropic future attractor that is reached sufficiently quickly by any trajectories.

It is clear that in general there are two types of solutions that reach the zero value of the average Hubble parameter $H=0$: bounce solutions (that we have already discussed) and re-collapse solutions that can appear during expansion. We try to find this kind of solutions numerically by using the following setup. We start numerical integration with the points with $H_0=0$ which are either definitely or bounce or re-collapse solutions and watch in which direction evolution occurs for different values $T_0$ and different parameters $w$, $f_0$. It is clear that there exist two ways from this kind of points: one leads to a singularity and therefore we can interpret it as re-collapse solutions and the other leads to a future isotropic attractor (and therefore bounce solutions). To stop numerical integration for bounce solutions, we used the criteria $|a-\frac{2}{3}|<10^{-4}$. We studied the following set of parameters: $f_2=\pm 0.1$, $w=\pm\frac{2}{3},\,\pm\frac{1}{3},\,0$ and scan the region $T_0\in[ 10^{-2},\,0.91 ]$ with the step $10^{-2}$. No any re-collapse solutions were detected.

\subsection{Full phase portraits}

In general, we have already described all possible stationary points of studying the system in sec.\ref{sec:5}, but usually it is not enough to get eigenvalues for stationary points to find the full dynamical picture; moreover, for some degenerate points this task is very cumbersome. For this reason we used the system of computer algebra to build full phase portraits and find both expected and unexpected results. We discuss here only the functions in the form $$f=T+f_NT^N,$$ and the solution for the point {\bf P2} from sec.\ref{sec:5} reads
\begin{equation}
T_0=\left[\frac{1-w}{\left( 2Nw -w -1 \right)f_N} \right]^{\frac{1}{N-1}},
\label{1.28}
\end{equation}
and it exists for \{even $N$, $[w(2N-1)-1]f_N<0$\} $\bigcup$  \{odd $N$, $[w(2N-1)-1]f_N>0$\};
whereas for the point {\bf P3} we have
\begin{equation}
T_0=\left[\frac{-1}{\left( 2N -1 \right)f_N} \right]^{\frac{1}{N-1}},
\label{1.29}
\end{equation}
and therefor it exists for  \{even $N$, $f_N>0$\} $\bigcup$  \{odd $N$, $f_N<0$\}.

Since the first point {\bf P1} always exists, we have four possibilities.

\begin{figure}[h]
  \includegraphics[width=10cm]{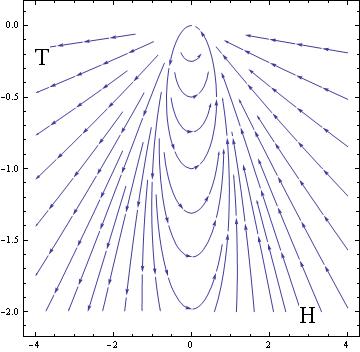}
  \caption{Phase portrait for $f_2=-0.2$, $N=2$, $w=0$, $\kappa^2=1$.}
  \label{fig:fig5}
\end{figure}

{\bf I}. The only point {\bf P1} exists. The typical picture for this case is presented in fig.\ref{fig:fig5}.

\begin{figure}[h]
  \includegraphics[width=10cm]{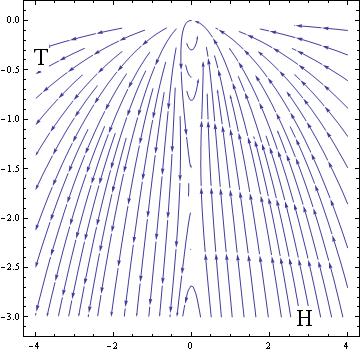}
  \caption{Phase portrait for $f_2=-0.2$, $N=2$, $w=\frac{2}{3}$, $\kappa^2=1$.}
  \label{fig:fig6}
\end{figure}

{\bf II}. The points {\bf P1} and {\bf P2} exist. The typical picture for this case is presented in fig.\ref{fig:fig6}.

\begin{figure}[h]
  \includegraphics[width=10cm]{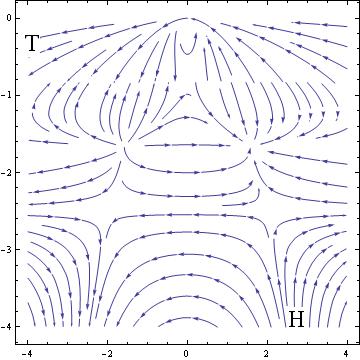}
  \caption{Phase portrait for $f_2=0.2$, $N=2$, $w=\frac{2}{3}$, $\kappa^2=1$.}
  \label{fig:fig7}
\end{figure}

{\bf III}. The points {\bf P1} and {\bf P3} exist. The typical picture for this case is presented in fig.\ref{fig:fig7}.

\begin{figure}[h]
  \includegraphics[width=10cm]{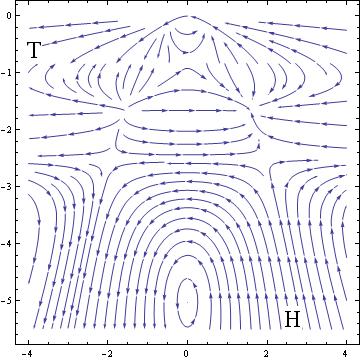}
  \caption{Phase portrait for $f_2=0.2$, $N=2$, $w=0$, $\kappa^2=1$.}
  \label{fig:fig8}
\end{figure}

{\bf IV}. All three types of points exist. The typical picture for this case is presented in fig.\ref{fig:fig8}.

\section{Discussion} \label{sec:7}

In this paper, we have studied the dynamic of the Bianchi I Universe filled with the perfect fluid in the teleparallel $f(T)$-gravity. We focused on power law functions and mainly quadratic, but some of our results are correct for any rational types of functions. We explicitly demonstrated that a future isotropic attractor is quite natural in this kind of theories and exists practically for any function and any matter and proved its stability using numerical research. Another important thing is the existence of a huge number of bouncing solutions, which appear for any matter and functions. We detected that for some specific values of parameters the dynamical picture became much richer: appear de Sitter points {\bf P3} and very specific point {\bf P2}. Simultaneously, with the appearance of de Sitter points we can see in fig.\ref{fig:fig7} and fig.\ref{fig:fig8} the emergence some specific horizontal line (name it {\bf L1}), which divides the visible region in two parts. To understand this phenomenon, let us go back to eq.(\ref{1.18}). We can see that the coefficient of the derivative term is $(2Tf''+f')$ and it is equal to zero for the quadratic function on the line $T_*=-1/6f_2$. Thus, we can see that this line is nothing more than the place where the coefficient of the highest derivative is equal to zero and trajectories cannot cross it.

Now let us discuss another interesting possibility. Up to this point, we have implied that $f_2\ll 1$, but if we violate this condition and put the contrary $f_2\gg 1$, the dynamical picture will change drastically. Indeed, if we put $f_2=1/2H_0^2$, where $H_0$ is the modern value of the Hubble parameter, the de Sitter point {\bf P3} will replaced to the physical region, it will still be stable and therefore can be used for describing the late time acceleration stage. It is to be noted that in this case the old future isotropic attractor will be cut off from the real dynamics by the line {\bf L1}. Fortunately, using additional numerical investigations we found that there appeared a new one which led to de Sitter point. In this case, phase portrait transforms from the typical picture in fig.\ref{fig:fig8} and becomes very similar to fig.\ref{fig:fig5}. Also, it should be noted that for this kind of scenario it is impossible to prevent the appearance of the point {\bf P2} for dust matter but only for some specific matter with $w(2N-1)-1>0$. It means that for the dust filled Universe the point {\bf P2} will exist as well and it is situated near the de Sitter point. As we can see from fig.\ref{fig:fig8}, the trajectories near this point have a circle form; thus we can expect a number of re-collapse solutions to appear.

It is to be noted that our research can not claim to be overall, because the main results concerning stability were obtained using a numerical approach and only for a quadratic function. Nevertheless, we detect a few general tendencies and demonstrate that it is possible to construct a viable cosmological scenarios for reasonable (from a physical point of view) values of parameters.
	
Finally, let us compare our results with the previous ones \cite{ST2} in more detail. First of all, note that the future isotropic attractor and de Sitter point were obtained there correctly; however, due to the use of dimensionless variables the physical meaning of some points is not so obvious. Moreover our point {\bf P2} was not found in that research. We found a combination of parameters $w(2N-1)-1$ as a condition for the existence of our point {\bf P2}, whereas in the cited paper it appears as a condition for isotropy of the initial singularity. It can be explained as follows. Our point {\bf P2} was missed in that research, whereas we did not investigate the structure of the initial singularity. And finally, we found a huge number of bounce solutions, whereas in that paper they are absent at all. The existence of bouncing solutions in the $f(T)$-gravity was well known for the FRW Universe with spatial curvature \cite{ST3,CSSZ}, but for the Bianchi I Universe this result is absolutely novel.

\begin{acknowledgments}
The author is grateful to A.V. Toporensky for discussion and some useful remarks.
This work was supported by the RFBR grant 20-02-00411 A.

\end{acknowledgments}


\begin{thebibliography}{99}
\bibitem{Einstein} A. Einstein,  Math. Annal. {\bf 102}, 685 (1930),  A. Unzicher, T. Case, arXiv:physics/0503046.

\bibitem{AP}
R. Aldrovandi and J.G. Pereira, Teleparallel Gravity:
An Introduction (Springer, Dordrecht, 2012)

\bibitem{JRSV}
Laur Jarv, Mihkel Runkla, Margus Saal, Ott Vilson, Nonmetricity formulation of general relativity and its scalar-tensor extension, Phys. Rev. D 97, 124025 (2018); arXiv:1802.00492 [gr-qc]

\bibitem{LSB}
Baojiu Li, Thomas P. Sotiriou, John D. Barrow, f(T) Gravity and local Lorentz invariance, Phys.Rev.D83:064035,2011;	arXiv:1010.1041 [gr-qc]

\bibitem{SLB}
Thomas P. Sotiriou, Baojiu Li, John D. Barrow, Generalizations of teleparallel gravity and local Lorentz symmetry, Phys.Rev.D83:104030,2011;  	 arXiv:1012.4039 [gr-qc]

\bibitem{TB}
Nicola Tamanini, Christian G. Boehmer, Good and bad tetrads in f(T) gravity, Physical Review D86 (2012) 044009;	arXiv:1204.4593 [gr-qc]

\bibitem{KS2}
Martin Krssak, Emmanuel N. Saridakis, The covariant formulation of f(T) gravity, Class. Quantum Grav. 33 (2016) 115009; arXiv:1510.08432

\bibitem{G2} Alexey Golovnev, Tomi Koivisto, Marit Sandstad, On the covariance of teleparallel gravity theories, Classical and Quantum Gravity 34 (2017) 145013, arXiv:1701.06271

\bibitem{HJU} Manuel Hohmann, Laur Jarv, Ulbossyn Ualikhanova, Covariant formulation of scalar-torsion gravity, Phys. Rev. D 97, 104011 (2018); arXiv:1801.05786

\bibitem{G3}
Alexey Golovnev, Introduction to teleparallel gravities, arXiv:1801.06929

\bibitem{Linder}  E. V. Linder, Phys. Rev. {\bf D 81}, 127301 (2010); arXiv:1005.3039 [astro-ph.CO]

\bibitem{KS1}
Georgios Kofinas, Emmanuel N. Saridakis, {\it Phys. Rev. D} {\bf 90}, 084044 (2014); arXiv:1404.2249
 	
\bibitem{Tret}
Petr V. Tretyakov, Dynamical stability of extended teleparallel gravity,  	Mod. Phys. Lett. A, Vol. 31, No. 14 (2016) 1650085; arXiv:1602.01287 [gr-qc]

\bibitem{arg1}  R. Ferraro and F. Fiorini, Phys. Lett. {\bf B 70}, 75 (2011)

\bibitem{arg2}
 R. Ferraro and F. Fiorini, Phys. Rev. {\bf D 84}, 083518 (2011)

\bibitem{BMNO}
Kazuharu Bamba, Ratbay Myrzakulov, Shin'ichi Nojiri, Sergei D. Odintsov, Reconstruction of f(T) gravity: Rip cosmology, finite-time future singularities and thermodynamics, Physical Review D 85, 104036 (2012); arXiv:1202.4057 [gr-qc]

\bibitem{BCNO}
Kazuharu Bamba, Salvatore Capozziello, Shin'ichi Nojiri, Sergei D. Odintsov, Dark energy cosmology: the equivalent description via different theoretical models and cosmography tests, Astrophysics and Space Science (2012) 342:155-228; arXiv:1205.3421 [gr-qc]

\bibitem{BOSG}
Kazuharu Bamba, Sergei D. Odintsov, Diego Saez-Gomez, Conformal symmetry and accelerating cosmology in teleparallel gravity, Physical Review D 88, 084042 (2013); arXiv:1308.5789 [gr-qc]

\bibitem{BCLNSG}
Kazuharu Bamba, Salvatore Capozziello, Mariafelicia De Laurentis, Shin'ichi Nojiri, Diego Saez-Gomez, No further gravitational wave modes in F(T) gravity, Physics Letters B 727 (2013) 194-198; arXiv:1309.2698 [gr-qc]

\bibitem{N1}
G. G. L. Nashed, Regularization of f(T) gravity theories and local Lorentz transformation, Advances in High Energy Physics, Volume 2015 (2015), Article ID 680457, 8 pages;	arXiv:1403.6937 [gr-qc]

\bibitem{CCLS}
Yi-Fu Cai, Salvatore Capozziello, Mariafelicia De Laurentis, Emmanuel N. Saridakis, f(T) teleparallel gravity and cosmology, Rept.Prog.Phys. 79 (2016) no.4, 106901; arXiv:1511.07586 [gr-qc]

\bibitem{N2}
A. Awad, W. El Hanafy, G.G.L. Nashed, Emmanuel N. Saridakis, Phase Portraits of general f(T) Cosmology, JCAP 1802 (2018) no.02, 052; arXiv:1710.10194 [gr-qc]

\bibitem{CLSX}
Yi-Fu Cai, Chunlong Li, Emmanuel N. Saridakis, LingQin Xue, f(T) gravity after GW170817 and GRB170817A, Phys. Rev. D 97, 103513 (2018), arXiv:1801.05827 [gr-qc]

\bibitem{WG}
Yi-Peng Wu, Chao-Qiang Geng, Matter Density Perturbations in Modified Teleparallel Theories, JHEP 2012, 142 (2012);	arXiv:1211.1778 [gr-qc]

\bibitem{G1}
Alexey Golovnev, Tomi Koivisto, Cosmological perturbations in modified teleparallel gravity models, JCAP 2018(11) 012 (2018); Report number: 	NORDITA 2018-073, DOI:10.1088/1475-7516/2018/11/012; arXiv:1808.05565 [gr-qc]

\bibitem{TT}
A.Toporensky, P.Tretyakov, Spin connection and cosmological perturbations in f(T) gravity, Phys. Rev. D 102, 044049 (2020); arXiv:1911.06064 [gr-qc]

\bibitem{SB1}
S. Bahamonde et al., Teleparallel Gravity: From Theory to Cosmology, arXiv:2106.13793 [gr-qc]

\bibitem{SS1}
M. Sharif, Shamaila Rani, F(T) Models within Bianchi Type I Universe, Mod.Phys.Lett.A 26 (2011) 1657; arXiv:1105.6228 [gr-qc]

\bibitem{RHSGR}
M. E. Rodrigues, M. J. S. Houndjo, D. Saez-Gomez, F. Rahaman, Anisotropic Universe Models in f(T) Gravity, Phys. Rev. D 86, 104059 (2012); arXiv:1209.4859 [gr-qc]

\bibitem{SA}
M. Sharif, Sehrish Azeem, Dark Energy Models and Cosmic Acceleration with Anisotropic Universe in f(T) Gravity, Commun.Theor.Phys. 61 (2014) 4, 482

\bibitem{N3}
G. G. L. Nashed, Exact homogenous anisotropic solution in f(T) gravity theory, Eur. Phys. J. Plus 129, 188 (2014)

\bibitem{SS2}
M. Sharif, Saima Jabbar, Phase Space Analysis and Anisotropic Universe Model in f(T) Gravity,  Commun.Theor.Phys. 63 (2015) 2, 168

\bibitem{PBL}
Andronikos Paliathanasis, John D. Barrow, P.G.L. Leach, Cosmological Solutions of f(T) Gravity, Phys. Rev. D 94, 023525 (2016); arXiv:1606.00659 [gr-qc]

\bibitem{WNI}
M.I. Wanas, G.G.L. Nashed, O.A. Ibrahim, Bianchi type I in f(T) gravitational theories, Chin.Phys.B 25 (2016) 5, 050401

\bibitem{PSB}
A. Paliathanasis, J. L. Said and J. D. Barrow, Stability of the Kasner Universe in f(T) Gravity, Phys. Rev. D 97, 044008 (2018); arXiv:1709.03432 [gr-qc]


\bibitem{ST1}
Maria A. Skugoreva, Alexey V. Toporensky, On Kasner solution in Bianchi I f(T) cosmology,  	Eur. Phys. J. C 78 (2018) no. 5, 377;  	arXiv:1711.07069 [gr-qc]

\bibitem{ST2}
Maria A. Skugoreva, Alexey V. Toporensky, Anisotropic cosmological dynamics in f(T) gravity in the presence of a perfect fluid, Eur. Phys. J. C 79 (2019) no. 10, 813; arXiv:1907.12538 [gr-qc]

\bibitem{LSB1}
Baojiu Li, Thomas P. Sotiriou, John D. Barrow, Large-scale Structure in f(T) Gravity, Phys. Rev. D 83, 104017 (2011);
arXiv:1103.2786

\bibitem{ST3}
Maria A. Skugoreva, Alexey V. Toporensky, Bouncing solutions in f(T) gravity, Eur. Phys. J. C 80 (2020) no. 11, 1054; arXiv:2009.11410 [gr-qc]

\bibitem{CSSZ}
Alessandro Casalino, Bruno Sanna, Lorenzo Sebastiani, Sergio Zerbini, Bounce Models within Teleparallel modified gravity, Phys. Rev. D 103, 023514 (2021); arXiv:2010.07609 [gr-qc]






\end{thebibliography}
\end{document}